# Defending against malicious peripherals with Cinch

Sebastian Angel,*† Riad S. Wahby,‡ Max Howald,§† Joshua B. Leners,∥
Michael Spilo,† Zhen Sun,† Andrew J. Blumberg,* and Michael Walfish†

*The University of Texas at Austin  †New York University  ‡Stanford University  §The Cooper Union  ∥Two Sigma

**Abstract**

Malicious peripherals designed to attack their host computers are a growing problem. Inexpensive and powerful peripherals that attach to plug-and-play buses have made such attacks easy to mount. Making matters worse, commodity operating systems lack coherent defenses, and users are often unaware of the scope of the problem. We present Cinch, a pragmatic response to this threat. Cinch uses virtualization to attach peripheral devices to a logically separate, untrusted machine, and includes an interposition layer between the untrusted machine and the protected one. This layer regulates interaction with devices according to user-configured policies. Cinch integrates with existing OSes, enforces policies that thwart real-world attacks, and has low overhead.

## 1 Introduction

Peripheral devices are now powerful, portable, and plentiful. For example, the inexpensive "conference USB sticks" that we have all received include not only the stored conference proceedings but also a complete computer. Given this trend, it is easy to create *malicious* peripheral devices [43, 61, 88, 98]. And yet, it is difficult to defend against them: commodity machines and operating systems continue to be designed to trust connected peripherals.

Consider a user who is induced to insert a malicious USB stick into his or her laptop [91, 135, 148]. There are now many examples [16, 75, 89] of such devices injecting malware (most infamously, Stuxnet [94]), by exploiting vulnerabilities in the host's drivers or system software.

Another alarming possibility is that, while following the USB specifications, the malicious device can masquerade as a keyboard. The device can then use its keystroke-producing ability to install a virus or exfiltrate files [43, 61, 125, 150]. As a last example, a USB device can eavesdrop on the communication between another device, such as the user's true keyboard, and the host [12, 17, 25, 72, 124].

These problems will get worse: on next-generation laptops [5, 10], *all* ports, including the power port, are USB, which means that any of the attacks above could be carried out by a malicious charger. For that matter, your phone might be compromised right now, if you borrowed a USB charger from the wrong person.

On the one hand, the concepts needed to solve these problems have long been understood. For example, in Rushby's separation kernel [129] (see also its modern descendants [81, 122]), the operating system is architected to make different resources of the computer interact with each other as if they were members of a distributed system. More generally, the rich literature on high-assurance kernels offers isolation, confinement, access control, and many other relevant ideas. On the other hand, applying these works in full requires redesigning the operating system and possibly also the hardware.

Solutions that target device security for today's commodity systems are not adequate for the task, often because they were designed under different models (§8). For example, work on device driver containment [80, 83, 93, 95, 96, 105, 112, 114, 127, 143–145, 152] and reliability [108, 130–132] trusts devices or assumes they are at worst buggy; the attacks mentioned earlier are largely out of scope. Hotplug control frameworks [13, 15, 18, 22, 33, 35, 37, 48, 50, 55], of which a notable example is udev on Linux [56, 110], enable users to express that certain devices should be denied access. However, access is all-or-nothing, decisions are based upon the device's claimed identity rather than its ongoing behavior, and a malicious device can disarm the enforcement mechanism. Qubes [45] protects the OS and applications from malicious USB devices, but achieves its strong guarantees at the expense of functionality.

The fundamental issue is that the I/O subsystems in commodity operating systems do not have an organizing abstraction that could serve as a natural foundation for security features. This paper attempts to fill that void.

Our point of departure is a simple suggestion: rather than design a new framework, *why not arrange for attached peripheral devices on commodity operating systems to appear to the kernel as if they were untrusted network endpoints?* This would create an interposition point that would allow users and administrators to defend the rest of the computer, just as firewalls and other network middleboxes defend hosts from untrusted remote hosts. Our animating hope is that a system based on this picture would eliminate large classes of vulnerabilities, be easy to deploy, and enable new functionality. To explore that vision, this paper describes the design, implementation, and experimental evaluation of a system called *Cinch*. Cinch begins with the following requirements:

- *Cinch should make peripheral buses look "remote," despite the physical coupling*, by preventing direct inter-

action with the rest of the computer (memory access, interrupts, etc.).

- *Under Cinch, traffic between the "remote" devices and the rest of the computer should travel through a narrow choke point.* This choke point then becomes a convenient location for deploying defenses that inspect and mediate interactions with untrusted devices.
- *Cinch should not require modifying bus standards, motherboards, OSes, or driver stacks.* Any of these would be massive undertakings, would have to be done for multiple platforms, and would jettison the immense effort behind today's installed base.
- *Cinch should be portable*, in the sense that Cinch itself should not need to be re-designed or re-implemented for different operating systems.
- *Cinch should be flexible and extensible*: users, operators, and administrators should be able to quickly develop and deploy a wide range of defenses.
- *Cinch should impose reasonable overhead* in latency and throughput.

Cinch responds to these requirements with the following architecture, focused on USB as a target (§4). Under Cinch, USB devices attach to an isolated and untrusted module; this is enforced via hardware support for virtualizing I/O [70, 71]. The untrusted module tunnels USB traffic to the protected machine, and this tunnel serves as a choke point for enforcing policy.

To showcase the architecture, we build several example defenses (§5). These include detecting attacks by matching against a database of attack signatures (§5.1); sanitizing inputs by ensuring that messages and device state transitions comply with protocol and device specifications (§5.2); sandboxing device functions and enforcing hotplug policies (§5.3); device authentication and traffic encryption (§5.4); and logging and remote auditing (§5.5).

Our implementation of Cinch (§6) instantiates both the untrusted module and the protected machine as separate virtual machines. As a consequence, Cinch protects any OS that runs atop the underlying hypervisor. In principle, these virtualization layers can be reduced or eliminated, at the cost of development effort and portability (§4.2).

To study Cinch's effectiveness, we developed exploits based on existing vulnerabilities [14], performed fuzzing, and conducted an exercise with a red team whose members were kept isolated from Cinch's development (§7.1–§7.3). Our conclusion is that Cinch can prevent many attacks with relatively little operator intervention. We also find that developing new defenses on Cinch is convenient (§7.4). Finally, Cinch's impact on performance is modest (§7.5): Cinch adds less than 3 milliseconds of latency and can handle USB 3 transfers of up to 2.1 Gbps, which is 38% less than the baseline of 3.4 Gbps.

Cinch is enabled—and inspired—by much prior work in peripherals management, hardware-assisted virtualization, privilege separation, and network security. We delve into this work in Section 8. For now, we simply state that although Cinch's individual elements are mostly borrowed, it is a novel synthesis. That is, its contributions are not mechanical but architectural. These contributions are: viewing peripherals as remote untrusted endpoints, and the architecture that results from this perspective; the instantiation of that architecture, which uses virtualization techniques to target a natural choke point in device driver stacks; a platform that allows defenses to existing attacks to be deployed naturally on commodity hardware, in contrast to the status quo; and the implementation and evaluation of Cinch.

Cinch is not perfect. First, it shrinks the attack surface that the protected machine exposes to devices, but introduces new trusted code elsewhere (§4.2). Second, although Cinch can reduce the universe of possible inputs to the drivers and OS on the protected machine (by ruling out noncompliant traffic), a malicious device might still exploit bugs in how the code handles *compliant* traffic. On the other hand, the user can decide which devices get this opportunity; further, addressing buggy drivers and system software is a complementary effort (§8). Third, Cinch does not unilaterally defend against higher-level threats (data exfiltration, malware, etc.); however, Cinch creates a platform by which one can borrow and deploy known responses from network security (§5). Finally, some of Cinch's defenses require changes within the device ecosystem (§9). For example, defending against masquerading attacks requires device (but not bus) modifications. However, these changes are limited: in our implementation, one person prototyped them in less than two days (§6.3). Importantly, these changes can be used with unmodified legacy devices via an inexpensive adapter.

Despite its shortcomings, Cinch is a substantial improvement over the status quo when considering the misbehavior that it rules out and the functionality that it enables. Moreover, we hope that Cinch's perspective on device security will be useful in its own right.

## 2 Background: Universal Serial Bus (USB)

Commodity computing devices (phones, tablets, laptops, workstations, etc.) have several peripheral buses for pluggable devices. These include USB [57, 58], Firewire [1], and Thunderbolt [54]. Cinch focuses on USB as an initial target; we make this choice because USB is ubiquitous and complex, and because it has become a popular locus of hardware-based attacks. However, our approach applies to other buses.

Figure 1 depicts the hardware and software architecture of USB. USB is a family of specifications for connecting and powering peripheral devices. Bandwidth ranges from 1.5 Mb/s (USB 1.0) to 10 Gb/s (USB 3.1). Example devices include storage (e.g., memory sticks), keyboards, sound cards, video cameras, Ethernet adapters, and smart card readers. These devices connect to a host (for example, a laptop or desktop). Some computers can act as either a device or a host; for example, a smart phone or laptop can appear as a storage device or power consumer to a desktop, but as a host to a keyboard.

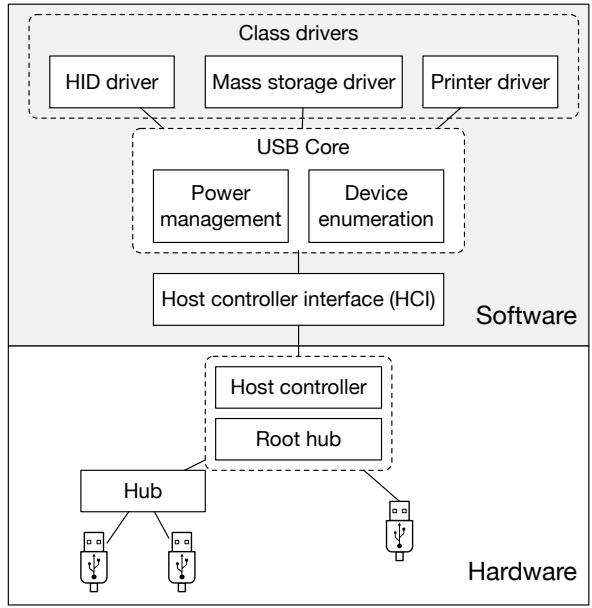

FIGURE 1—The hardware and software of a USB stack (§2). Both physical devices and drivers are arranged hierarchically; devices are rooted at the host controller, and drivers are rooted at the host controller interface. Components in dashed boxes are logically in the same layer of the USB stack.

**USB hardware.** USB has a tree topology. Each device has an *upstream* connection to a *hub*. Hubs multiplex communication from one or more *downstream* devices, and are themselves devices with an upstream connection to another hub or to the root of the tree. The root is a *host controller*, which connects to the host by, for example, PCIe. The host controller acts as the bus master: it initiates all transfers to and from devices, and devices are not permitted to transmit except when polled by the host controller. Also, the host controller issues interrupts to the host and has direct access to host memory via DMA.

**USB protocol.** The USB specifications [57, 58] define a protocol stack comprising three layers. The bottom layer includes electrical specifications and a low-level packet protocol. The middle layer of the stack includes addressing, power management primitives, and high-level communication abstractions. USB devices, comprising one or more *functions*, sit at the top of the stack. Functions act as logically separate peripherals that are exposed by a single physical device. For example, a phone might expose a camera function, a network adapter function, and a storage function. Each of these functions is associated with its own high-level driver software.

**USB driver architecture.** The USB specification describes three layers of software abstraction on the host. The lowest level, the *host controller interface* or *HCI*, configures and interacts with the host controller hardware via a local bus (e.g., PCIe). An HCI driver is particular to a host controller's hardware interface but exposes a hardware-independent abstraction to the next software layer, called *core*. Core manages device addressing and power management, and exposes an interface for high-level drivers to communicate with devices. Core also *enumerates* devices when they are attached, which entails identifying the device and activating its driver.

The uppermost layer, *class drivers*, are high-level drivers that interact with functions (as described above). These drivers provide an interface between USB devices and the rest of the OS. For example, a keyboard's class driver interacts with the kernel's input subsystem. Another example is the mass storage class driver, which talks to the kernel's storage subsystem. The USB specification defines a set of generic classes for a broad range of devices, e.g., keyboards, mice, network interfaces, storage, cameras, audio, and more. Operating systems generally include support for a large subset of the generic classes, allowing devices to leverage preexisting drivers.

## 3 Causes, threat model, and taxonomy

### 3.1 Why is USB so vulnerable?

The root of the problem is the implicit assumption that hardware is inherently trustworthy, or at worst buggy but non-malicious. As a consequence, neither USB nor mainstream OSes are designed to be robust in the face of malicious devices. One manifestation of this is the lack of authentication or confidentiality guarantees at any layer of the USB standard. As examples, devices self-report their identity and capabilities without authentication; the communication primitives at all layers of the protocol stack (§2) are cleartext; and, prior to USB 3, host-to-device messages are broadcast across the entire bus [124].

A related issue is that the USB protocol and common driver stacks emphasize convenience above correctness and security. For example, hotplugged devices are often activated without user confirmation. Coupled with the lack of device authentication, this means that the OS cannot determine what device the user intended to connect, or even that a hotplug event was generated by the *user* rather than a malicious device [24, 75]. Moreover, malicious device makers can rely on the near universal availability of generic class drivers (e.g., for keyboards), since users expect these devices to "just work."

The range and sophistication of USB-based threats has escalated substantially in recent years. Whereas hardware design costs were once a barrier to entry, creating custom USB devices is now cheap, both in dollars and development time [43, 52, 61, 98, 100]; in fact, today's commodity USB devices are essentially software defined [43, 75, 98].

The press plays a role too: demonstrating USB attacks has become fashionable (e.g., recent media hype [6–8, 39, 53, 118] surrounding USB devices with reprogrammable firmware [43, 84, 125]). A third factor is ease of transmission: malicious USB devices can easily find their way into the hands of victims [148]. This is partly due to vulnerabilities in the supply chain [38, 74, 101, 141], such as adversarial manufacturers [102]. Intelligence agencies have also been known to use their resources to intercept and "enhance" shipments [27, 97], including conference giveaways [20, 21].

### 3.2 Threat model

We assume that devices can deviate from the USB specification arbitrarily. They may also violate the user's expectations, for example by masquerading as other devices or passively intercepting bus traffic. Alternatively, devices can present a higher-level threat; for example, a storage device can contain an invalid filesystem that triggers a bug in a filesystem driver. However, devices that cause physical damage to the host, with high voltage [86] for example, are out of scope.

We assume that the host's OS and drivers can be buggy but not malicious. We assume the same for the host's hardware besides the USB controller and USB devices.

### 3.3 A taxonomy of USB attacks

**Attacks on USB drivers.** USB drivers present an attack surface to devices. For example, a driver with an unchecked buffer access might allow a malicious device to overwrite kernel memory via an overflow. The space of possible misbehavior here is vast. For instance, devices might try to deliver more data to the driver than indicated by the device's configuration [31]; claim impossible configurations [28, 30]; exceed limits prescribed by USB class specifications [4, 32, 42]; or produce otherwise invalid or nonsensical reports [75, 87–89, 92, 137].

The prevalence of these attacks reflects a difficult software engineering situation. Since a driver writer needs to be prepared for an enormous range of undocumented behavior, drivers need lots of error checking code; such code is often ill-exercised and creates complexity, leading to more vulnerabilities. Indeed, more than half of the vulnerabilities related to USB drivers in the CVE database [14] are the result of improper handling of noncompliant USB transfers; many more such vulnerabilities likely remain undisclosed [87, 137].

**Other attacks on the host via USB.** USB can also expose the rest of the host system's kernel or user software to attacks by malicious devices. Recall that USB class drivers provide an interface between devices and other kernel subsystems (§2). Leveraging this interface, a USB flash drive might be used to attack the kernel's storage or filesystem drivers [19, 44, 63, 64]. Or the drive might carry a virus [94] or covertly steal data [140].

Of particular concern is the possibility of attacks in which the USB host controller uses DMA (direct memory access) to bypass the CPU and read or write arbitrarily to RAM [26, 116, 139, 142]. A successful DMA attack neutralizes essentially all software security measures, allowing the attacker to steal sensitive data, modify running software (including the kernel itself), and execute arbitrary code [128]. And the host controller does not need to be malicious: misconfigured DMA-capable hardware is a proven vector for such attacks [153, 154].

**Privacy and authentication threats.**

*Device masquerading.* When a device is plugged in, the host asks the device for information about its capabilities. The device can respond, disguised as another device or even another class [29, 41, 43, 65, 85, 120, 125, 150]. For example, Psychson [43] enables rewriting the firmware on a cheap USB storage device so that it will act like a keyboard; similarly, the commercially available "USB Rubber Ducky" [61] is a programmable keystroke injector in the guise of a flash drive. Likewise, a malicious hub can masquerade as other devices [25]. These examples are more than idle threats: penetration testers regularly use such tools to breach security systems [3, 24].

*Bus eavesdropping.* In USB 2 and earlier versions, hubs broadcast traffic from their upstream port to all downstream ports (§2), so any device on the bus can eavesdrop on traffic from the host to any other device [124]. In all protocol versions, malicious hubs can eavesdrop on upstream and downstream traffic [17, 72]. Furthermore, a hub need not be malicious: if its firmware is buggy, it can be exploited by a malicious device [25].

## 4 Architecture and rationale

The top-level goal of Cinch is to enforce security policies that enable safe interactions between devices and the host machine. This enforcement must be done in a way that respects the requirements outlined in Section 1. In particular, we must answer two questions in the context of USB: (1) Where and how can one create a logical separation between the bus and the host, while arranging for an explicit communication channel that a policy enforcement mechanism can interpose on? (2) How can one instantiate this separation and channel with no modifications to bus standards, OSes, or driver stacks?

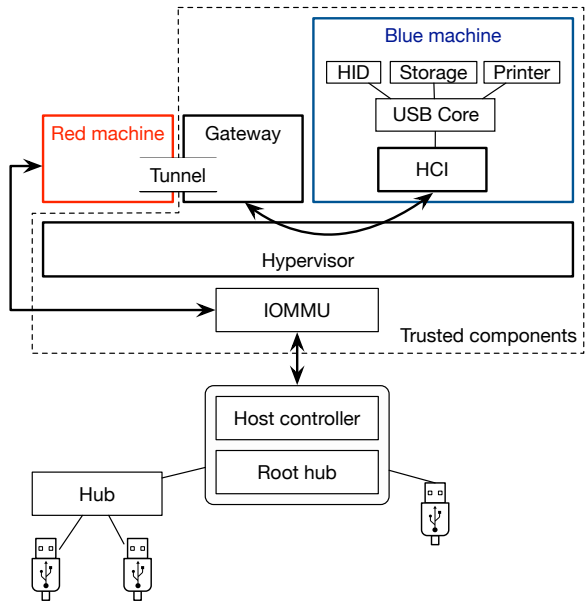

FIGURE 2—The architecture of Cinch. The trusted components are surrounded by the dashed line. I/O virtualization separates the USB host controller from the blue machine's HCI, redirecting DMA and interrupts to the red machine. The red machine encapsulates and sends USB transfers through the Tunnel to the Gateway. Once the Gateway has applied all security policies, it redirects those transfers to the blue machine's HCI.

We begin with the logical separation, which Cinch enforces at the boundary between the host controller and its driver (HCI), depicted in Figure 1. We choose this separation point for two reasons: first, it results in a narrow choke point where software can interpose. Second, the host controller is "dangerous"—it issues interrupts and accesses memory via DMA (§2, §3.3)—so there should be a barrier between it and the rest of the system, including the modules that administer policy decisions.

The architecture is depicted in Figure 2. After logically separating the host controller, Cinch attaches it to a new module, the *red machine*. The red machine is an endpoint to a communication channel, the *Tunnel*. The other endpoint, the *Gateway*, is positioned at the entrance to the host that Cinch protects, the *blue machine*. (These names are inspired by Lampson's red/green machine partitioning [111].) The Gateway mediates all transfers through the Tunnel and enforces security policies (for example, dropping or rewriting USB traffic, as described in §5) before those transfers reach the blue machine's USB stack.

To connect the host controller to the red machine, Cinch uses I/O virtualization hardware, which is widely available in modern CPUs [70, 71]. Specifically, an IOMMU provides address translation and protection, which restricts a physical device's DMA transfers to a designated memory region (in this case, that of the red machine); and interrupt remapping provides analogous translation and protection for interrupts.

### 4.1 Instantiation

In our current implementation, the lowest layer of software—the one that manages the hardware resources and configures the I/O virtualization hardware—is a combination of hypervisor and OS, and is trusted. The red machine runs on top of this hypervisor and is a full-fledged virtual machine, with a normal OS that has a stripped-down USB stack (§6.1). The blue machine is also a full-fledged virtual machine atop the hypervisor, and the Gateway is a separate process.

### 4.2 Discussion

With the instantiation described immediately above, Cinch meets the requirements described in Section 1. It isolates devices in the red machine, and its Gateway is a narrow choke point. It limits overhead to reasonable factors (§7.5), in part by leveraging hardware-assisted processor and memory virtualization [68, 123] (as distinct from I/O virtualization). It works with existing USB stacks; the main component needed is a driver in the hypervisor, to receive transfers from the Gateway. It works with a range of OSes because the blue machine runs unmodified. For the remaining requirements, flexibility is demonstrated in the next section (§5), and extensibility arises from Cinch's software structure (§6.2).

But a disadvantage is the size of the trusted computing base (TCB) and attack surfaces. Specifically, the TCB includes a full-featured hypervisor. The attack surface includes the red machine, which is running a full OS and which, if compromised, can attack the hypervisor and the blue machine via the virtualization interface (by attempting VM escapes, side channel inference, etc.).

There are a number of alternatives that, by tailoring the hypervisor and red machine, reduce the TCB at the cost of portability and additional development effort. As an extreme example, the blue machine could run directly on the host's hardware ("bare metal"), with the red machine running in an untrusted user-level process; the Gateway would also run in user space. In this setup, there would be no separate hypervisor; the blue machine would perform the few required hypervisor-like functions, such as configuring the I/O virtualization hardware to connect the host controller to the red machine process (see [77, 78, 126]). Compared to Cinch's instantiation, this one has a smaller TCB; it also has lower overhead, owing to the absence of virtual machines. However, it is less portable: each new blue machine OS needs corresponding "hypervisor" module and red machine implementations.

One can go further: the Gateway could entirely bypass the blue machine's USB stack, sending device traffic directly to the corresponding kernel subsystem (for example, sending USB keyboard events to the input subsystem). This would further reduce the TCB, at the cost of even more development work and less portability.

Another design point is a hardware-only solution: the red machine and Gateway would run on a device placed between USB devices and the blue machine, which would run as a normal, unmodified host. Compared to Cinch, this solution is potentially more portable, in that no software modifications or reconfiguration are needed. Further, this solution does not rely on I/O virtualization (which is widespread but not universal), and it leaves the host's virtualization hardware available for other uses. The disadvantages are that a hardware solution is likely to be less flexible, and that building hardware may be substantially more effort than building Cinch.

A non-solution, in our view, is to implement the Gateway in the host's USB stack, without a separate red machine. This setup does not have the separation discussed earlier; it would leave the host and Gateway vulnerable to DMA attacks by a compromised host controller.

# 5 Building defenses with Cinch

This section describes some of the defenses (which we call *Policies*) that Cinch supports, and the threats (§3) against which they defend. These Policies are not new; we discuss previous implementations in Section 8. The novelty is in providing a platform that makes a range of Policies straightforward to develop and deploy.

## 5.1 Detecting attacks by signature

The first strategy is signature matching: dropping messages that match a known pattern. Defenses in this class protect against attacks on drivers and user software (§3.3). The same strategy is used in network security (intrusion detection [47]) and desktop security (antivirus [11]) and has been effective in practice, as a first-line defense. The advantages and disadvantages hold in our context; we review them briefly.

To begin with, signature generation is flexible and can be done by victimized companies, individual users, and designated experts, based on observations of past attacks and reverse engineering of malicious devices. Further, shared databases of observed attack signatures can immunize others. This strategy also enables rapid responses to emerging threats: a signature of an attack is typically available long before the vulnerability is patched.

The principal disadvantage, of course, is that signatures generally provide protection only against previously observed attacks. Furthermore, they suffer from both false positives and false negatives: signatures that are too general may disable benign devices, while signatures that are too specialized can fail to catch all variants of an attack.

**Cinch's signature Policy.** We implement a signature matching module in Cinch that compares all USB traffic from the red machine to a database of malicious payload signatures. When a match occurs, Cinch disallows further traffic between the offending device and the blue machine.

## 5.2 Sanitizing inputs

Another class of defensive strategies detects when devices deviate from their specification; this is useful for defending against attacks on USB drivers (§3.3). Given a specification (say, provided by the manufacturer or converted from a standards document), Cinch checks that messages are properly formatted and that devices respond correctly to commands. While drivers can (and in some cases, do) implement such checks, moving enforcement to a dedicated module can eliminate redundant code and reduce driver complexity (§3.3, "Attacks on USB drivers").

A related strategy in Cinch modifies apparent device behavior, either forcing adherence to a *strict subset* of the USB spec in order to match driver expectations, or else *relaxing* the USB spec by recognizing and fixing device "quirks"—behavior that is noncompliant but known to be benign—so that drivers need not do so.[1] This is closely related to traffic normalization [103], in which a firewall converts traffic to a canonical representation to aid analysis and ensure that decisions are consistent with end-host protocol implementations.

**Cinch's compliance Policy.** This Policy enforces device compliance with USB specifications. To build it, we manually processed the USB 2 and 3 specifications [57, 58], along with the specifications of five device classes (mass storage, HID, printer, power, and debug) [59]. The result is a module that monitors device states and transitions, and enforces invariants on individual messages and entire transactions. As a simple example, the compliance Policy checks that device-supplied identification strings are well formed—that is, that they comprise a valid UTF-16 string of acceptable length—and rewrites noncompliant strings. More complex state and transition checking is effected by keeping persistent information about each device for the duration of its connection.

Cinch's compliance Policy is conservative in handling noncompliance: if it cannot easily fix a device's behavior (for example, by rewriting identification strings as described above), it assumes the device is malicious and disables it.

**Cinch's assertion Policy.** This Policy implements the aforementioned relaxations and restrictions, by modifying how Cinch's compliance Policy regulates specific devices. As examples, a user might specify that a particular device's requests should be rewritten to work around buggy firmware. Or Cinch can require that devices handled by a certain driver must expose an interface that matches a specified template, obviating bug-prone compatibility checks in the driver's code (§3.3).

---

[1]In practice, many non-malicious devices fail to comply with the specification: the word "quirk" appears about once every 300 lines throughout the 300 kLoC Linux USB stack (!), and nearly all the devices we tested deviated from prescribed behavior in at least a small way.

## 5.3 Containing devices

This category includes querying a user for information about a newly connected device, restricting a device to a subset of its functionality, and isolating devices in private protection domains. Such defenses, which are useful against attacks on driver and user software and can foil masquerading attacks (§3.3), are forms of *hotplug control*. They decide—say, by asking the user—whether a newly connected device should be allowed to communicate with the blue machine, and if so, what functionality should be allowed. For example, Cinch might ask the user, "I see you just connected a keyboard. Is this right?"

In practice, such decisions can be much more complex. Recall from Section 2 that devices can define multiple functions, each of which is a logically separate peripheral. A careful user wishing to tether his or her laptop to a friend's phone could be informed of available functionality upon device connection, and choose to disallow the phone's storage function as a precaution against viruses.

Alternatively, the user might choose to connect the phone's file storage function to a separate protection domain—a sandbox—with limited capabilities and a narrow interface to the blue machine. In this case, the sandbox could scan files for viruses, and could expose a high-level interface (e.g., an HTTP or NFS server) to the blue machine. This approach leverages existing software designed for interacting with untrusted machines (in this case, a web or file browser), and can bypass many layers of software in the blue machine; on the other hand, it changes the interface to the device.

**Cinch's containment Policy.** We implement a "surgical" hotplug Policy: individual device functions can be allowed or disallowed, and the blue machine never interacts with disallowed devices. Cinch's Gateway can also sandbox whole devices or individual functions by redirecting selected USB traffic to separate protection domains that expose functionality to the blue machine through narrow interfaces, as described above.

## 5.4 Encryption and authentication

To handle devices that eavesdrop on the bus or masquerade as other devices, Cinch adapts well-known responses—authentication and encryption—to USB. For example, a user can disallow all keyboards except those having a certificate signed by a particular manufacturer. This prevents a malicious device without such a certificate from acting as a keyboard.

In more detail, a device authenticates to the Gateway by leveraging a trust relationship. As examples, manufactures sign certificates and install them on devices, and users are required to use devices whose certificates are signed by a trusted manufacturer; or users follow a pairing procedure as in Bluetooth [67] or GoodUSB [146], obviating a trusted manufacturer but adding a setup step. After completing a key exchange, the device and host share an encryption key. The user can then prevent masquerading and eavesdropping by installing a policy that disallows unauthenticated, untrusted, or unencrypted devices.

This arrangement raises several potential concerns: development overhead to build new devices, computational overhead for cryptography, and deployment on legacy devices. Below, we describe a proof-of-concept design that addresses these concerns. At a high level, the concerns are addressed by, respectively, abundant support for rapid development of embedded cryptographic applications [34, 36, 66], the speed of modern embedded processors, and a physical *adapter* that adds cryptographic functionality to legacy devices.

**Proof-of-concept USB crypto support.** To support authentication and encryption, we designed a *cryptographic overlay protocol*. This mechanism allows compatible devices to communicate with the Gateway via a TLS session that encapsulates all of their USB transfers.

To evaluate the crypto overlay, we built a *crypto adapter*, a physical device that sits between unmodified legacy devices and a host system running Cinch. The crypto adapter acts as a USB host for the legacy device, encapsulating and decapsulating the device's USB traffic inside a TLS session. To communicate this TLS-encrypted traffic to the host system, the crypto adapter also acts as a USB device attached to the host system, as we detail below. We refer to the crypto adapter's USB connection to the legacy device as the "inner" connection, and its connection to the host as the "outer" connection.

Two issues arise in designing the crypto overlay and adapter. First, a TLS session requires a full duplex stream transport, while USB's communication primitives are based on host-initiated polling (§2). This means that the outer USB connection cannot directly encapsulate a TLS session. Second, the Gateway does not implement a USB stack, meaning that, on its own, it cannot communicate with the crypto adapter via the outer USB connection.

To solve the first issue, Cinch uses an existing USB class that exposes a full-duplex Ethernet interface [59]; this Ethernet-over-USB traffic is carried by the outer USB connection. Then Cinch uses TCP over this Ethernet connection as the stream abstraction for TLS, yielding an indirect encapsulation of TLS in the outer USB connection.[2] To solve the second issue, we observe that, with the foregoing encapsulation, the Gateway need not handle the outer USB connection. Instead, the red machine treats the outer USB connection as an Ethernet device (thereby terminating the outer USB connection), and it forwards all packets it receives from that device to the Gateway via the Tunnel. Meanwhile, these packets are just the TCP stream carrying the TLS session, and thus the Gateway can talk TLS to the crypto adapter without a USB stack.

---

[2]An alternate approach with less overhead than TCP-over-IP-over-Ethernet-over-USB is to create a custom USB class providing a full-duplex stream abstraction with less generality than Ethernet.

Note that this arrangement differs from the way that Cinch handles other USB devices. For unencrypted devices, the Gateway receives USB transfers captured by the red machine; it inspects these transfers and then forwards them to the blue machine's HCI. But here, the Gateway receives packets (which the red machine decapsulated) that contain a TLS session. The Gateway decrypts to recover USB transfers, which it inspects and forwards.

**Cinch's crypto Policy.** Given devices implementing the crypto overlay, Cinch can enforce policies that rule out eavesdropping and masquerading by requiring authenticated and encrypted devices, as described at the outset of this section.

### 5.5 Logging and auditing

Logging is part of many defensive strategies: auditing logs can reveal anomalous behavior that might indicate a new attack. Moreover, logs can be used to develop new signature-based defenses (§5.1).

**Cinch's logging Policy.** Cinch's Gateway can be configured to log some or all traffic to and from the blue machine. Cinch can also replay logged data; we used this functionality to help develop attack signatures for our security evaluation (§7.3). Furthermore, Cinch can be configured to log to a remote server. This feature could allow real-time analysis of data from many different blue machines, for example in a corporate environment.

### 5.6 Extensions

Cinch enables usage scenarios beyond the ones described above. One example is data exfiltration prevention, which is often employed at the network level to address the threat of data theft [104, 115, 117, 133, 134], but is generally considered a more difficult problem in the context of USB [140]. By combining real-time remote auditing (§5.5) with signature detection (§5.1), Cinch allows administrators to apply exfiltration prevention policies to USB devices.

## 6 Implementation

We describe the components and the communication paths in our implementation of Cinch (§6.1). We also discuss the Policies implemented in Cinch, utilities that we use to create and test new exploits, and our method for deriving payload signatures (§6.2). Finally, we describe the proof-of-concept crypto adapter (§5.4) that we use to transparently provide encryption and authentication for existing USB devices (§6.3).

### 6.1 Components and communication paths

The hypervisor (§4.1) is Linux with KVM, meaning that virtual machines run in QEMU processes that are accelerated with virtualization hardware [68, 123]. In particular, Cinch requires hardware support for I/O virtualization [70, 71]. We tested with Intel hardware, but KVM also supports equivalent functionality from AMD.

The red machine runs Linux. It is configured to load only the HCI and core drivers (§2); higher-level USB drivers are not needed to capture USB transfers from devices. (An exception is the case of the crypto overlay, which requires a USB network driver; §5.4). The blue machine is another VM and, as stated in Section 4.2, can be any OS supported by QEMU. The Gateway runs as a user-level process on the Linux-KVM hypervisor.

The Tunnel between the red machine and the Gateway appears to both entities as a network device. The appeal of this approach is that the Tunnel connects to the untrusted part of the system (Figure 2, §4), and meanwhile IP stacks have been hardened over decades. Furthermore, this lets us leverage existing software for remotely accessing USB devices over a network [60, 73, 106]. Our implementation uses usbredir [73], which (on the red machine), captures USB transfers, listens on a network socket, and uses a custom protocol to encapsulate USB transfers inside a TCP stream.

As a usbredir client, the Gateway receives usbredir packets, filters or modifies them, and then, playing the role of a usbredir server, delivers them to the QEMU process running the blue machine. A module in QEMU is the corresponding client; it decapsulates the USB transfers (using usbredir) and injects them into a virtual host controller created by QEMU and exposed to the blue machine. From the virtual host controller, the USB transfers travel into the blue machine's HCI, with no software modifications on the blue machine.

Our implementation of Cinch supports USB versions through USB 3.

### 6.2 Gateway details

The Gateway is implemented in Rust [46]; it comprises about 8 kSLoC. Its major modules are parsers for usbredir packets and USB transfers, and a library that provides abstractions for creating new Policies. This library is inspired by the Click modular router [109] and provides domain-specific abstractions for USB (as examples, demultiplexing usbredir packets into USB transfers and filtering those transfers). As in Click, the user organizes modules into chains where one module's output is the next module's input. Several such chains can be configured to operate in parallel. Users configure module chains with files in JSON format.

| OS | exploit identifier | exploit description | prevention mechanism |
|---|---|---|---|
| **Windows 8.1** | 01:01:00:C:4 | Audio device with non-existent streaming interface | Signature Policy⋆ |
| | 01:01:00:C:5 | Audio device with invalid streaming interface | Signature Policy⋆ |
| | 03:00:00:C:16 | HID device with invalid report usage page | Compliance Policy |
| | 03:00:00:C:17 | HID device with invalid report usage page | Compliance Policy |
| | 09:00:00:C:9 | Hub with invalid number of ports | Compliance Policy |
| **Linux 4.2.0** | CVE-2016-2184 | Sound device with non-existent endpoint | Assertion Policy |
| | CVE-2016-2185 | RF remote control device with invalid interface or endpoint | Assertion Policy |
| | CVE-2016-2186 | Multimedia control device with invalid endpoint | Assertion Policy |
| | CVE-2016-2187 | Digitizer tablet device with invalid endpoint | Assertion Policy |
| | CVE-2016-2188 | I/O Warrior device with invalid endpoint | Assertion Policy |
| | CVE-2016-2384 | Audio device with invalid USB descriptor | Assertion Policy |
| | CVE-2016-2782 | Serial device with no bulk-in or interrupt-in endpoint | Assertion Policy |
| | CVE-2016-3136 | Serial device without two interrupt-in endpoints | Assertion Policy |
| | CVE-2016-3137 | Serial device without both in and out interrupt endpoints | Assertion Policy |
| | CVE-2016-3138 | Communication device without both control and data endpoints | Assertion Policy |
| | CVE-2016-3139 | Drawing tablet with invalid USB descriptor | Assertion Policy |
| | CVE-2016-3140 | Serial converter device with invalid USB descriptor | Assertion Policy |
| | CVE-2016-3951 | Communication device with invalid descriptor and payload | Compliance Policy |

⋆Exploit can be prevented with the compliance Policy, but we have not yet incorporated the necessary class specification (Audio) into Cinch.

FIGURE 3—Exploits for known-signature exercise (§7.1). Windows exploits were found by Boteanu and Fowler [79] with *umap* [88]; the reported identifier can be passed to umap using the "-s" flag to reproduce the exploit. We implemented the Linux exploits to target all USB-related CVEs from January–June 2016. The last column describes which Policy (§5) of Cinch prevents the exploit.

### 6.3 Proof-of-concept USB crypto adapter

We implement the crypto adapter (§5.4) using a Beagle-Bone Black [9] single-board computer that has a 1 GHz ARM Cortex-A8 processor and 512 MB RAM. For authentication, we generate a CA certificate and install it on the Gateway and crypto adapter. We use that CA certificate to sign certificates for the Gateway and crypto adapter, which mutually authenticate during the TLS handshake. The crypto adapter runs a version of usbredir that we augmented with support for TLS 1.2 [90] using OpenSSL [40]; these changes comprise less than 200 lines of code. The Gateway's crypto module uses stunnel [49] to listen for TLS connections.

## 7 Evaluation

Our evaluation of Cinch answers the following questions:

- *How effectively does Cinch defend against attacks?* We subject Cinch to known exploits (§7.1), fuzzing (§7.2), and a red team exercise (§7.3).
- *Can new functionality be developed and deployed on Cinch with ease?* We answer this question qualitatively, by relating our experiences (§7.4).
- *What is Cinch's performance overhead?* We examine latency and throughput (§7.5).

**Experimental hardware and OSes.** All of our experiments run on a single machine with a 3.3 GHz Intel i5-4590 and 16 GB of RAM. The hypervisor is Debian Jessie running Linux 4.2.0 with KVM enabled. The red machine's OS is also Debian Jessie running Linux 4.2.0. The blue machine's OS depends on the experiment and is either Windows 7 Ultimate SP1 (build 7601), Windows 8.1 Professional (build 9600), Debian Jessie with Linux 4.2.0, or Ubuntu 14.04 with a modified 4.2.0 kernel.

### 7.1 Known-signature attacks

We begin our evaluation of Cinch by subjecting it to synthetic attacks, based on documented vulnerabilities. For the attacks that succeed, we specify a "rematch" protocol, in which the operator can install a signature (§5.1) and then retry. This exercise is intended to address a counterfactual hypothetical: if Cinch had been deployed at the time of these vulnerabilities, would it have protected against their exploitation? And, if not, would a subsequent defensive reaction have been effective?

**Method and experiment.** We filter the CVE database [14] to select all the USB-related vulnerabilities reported from January to June of 2016. The resulting 13 CVEs apply to Linux 4.5 and earlier. For each CVE, we construct a payload that exploits it. We also include five exploits, disclosed by Boteanu and Fowler [79], that affect the most recent version of Windows 8.1; the targeted vulnerabilities are not in the CVE database.

Figure 3 summarizes the exploits. We confirm that each exploit successfully compromises the blue machine (Debian Jessie with Linux 4.2.0 or Windows 8.1) in the absence of Cinch. Once Cinch is enabled, we consider an attack successful if it compromises either the blue machine's kernel or the Gateway.

On the offensive side, we mount the attacks using a Facedancer [98]—a custom USB microcontroller that can masquerade as any USB device and issue arbitrary payloads when connected to the target machine. We program

|  | exploits prevented | |
| --- | --- | --- |
|  | match phase | rematch phase |
| **Known exploits (§7.1)** | | |
| Windows 8.1 | 3 / 5 | 5 / 5 |
| Linux 4.2.0 | 13 / 13 | 13 / 13 |
| **vUSBf [136, 137] payloads (§7.2)** | | |
| randomized devices | 10,000 / 10,000 | N/A |
| sample exploits | 13 / 13 | N/A |
| **red team round 1 (§7.3)** | | |
| Windows 7 | 2 / 2 | 2 / 2 |
| Linux 4.2.0 | 3 / 5 | 5 / 5 |
| **red team round 2 (§7.3)** | | |
| Windows 7 | 3 / 3 | 3 / 3 |
| Linux 4.2.0 | 11 / 16 | 13 / 16 |
| **red team round 3 (§7.3)** | | |
| Windows 7 | 3 / 3 | 3 / 3 |
| Linux 4.2.0 | 15 / 20 | 16 / 20 |

FIGURE 4—Summary of Cinch's security evaluation.

and control the Facedancer through a Python interface, using the GoodFET [99] and umap [88] tools.

On the defensive side, we configure Cinch with the signature, assertion, compliance, and logging Policies (§5). For the assertion Policy, we install 12 driver-specific configuration restrictions; these fix buggy or nonexistent checks, identified by the CVEs. For the signature Policy, we start with an empty signature database and check whether each attack succeeds; if it does, we craft a signature based on the payload and associated metadata, then conduct a rematch.

**Results** are summarized in Figure 4 ("Known exploits"); for each exploit, the mechanism that prevented it is listed in Figure 3. Cinch successfully detects and drops 16 offending payloads with no additional configuration. Two of the payloads were successful on their first try, but were blocked in the rematch phase; these payloads targeted vulnerabilities in the USB Audio class, which we have not yet included in Cinch's compliance Policy.

### 7.2 Fuzzing

Next, we assess the robustness of Cinch's compliance Policy (§5.2), via fuzz testing. We limit this exercise to attacks that target device enumeration, as implemented in the core and class drivers (§2). On the one hand, this is not a comprehensive exercise. At the same time, device enumeration is a common and well-studied source of vulnerabilities [137], accounting for about half of all USB-related entries in the CVE database.

In enumerating devices, USB core processes each device's USB *descriptors*: records, generated by the device, that identify its manufacturer, function, USB version, capabilities, etc. This process is complex because of the wide range of possible device configurations. Furthermore, the attack surface includes class driver initialization functions, since USB core passes descriptors to those functions; Schumilo et al. [137] demonstrate that many OSes and drivers do not handle device enumeration properly, especially when the device information is inconsistent or maliciously crafted.

**Method and experiment.** On the offensive side, we use vUSBf [136], a fuzzing tool that generates a random set of device descriptors and then emulates a device attach event. We update vUSBf to work with the most recent version of usbredir (v0.7.1), and we replace the red machine with an instance of vUSBf (that is, vUSBf communicates directly with the Gateway). In this setup, vUSBf can emulate hundreds of randomized devices per minute.

We run two experiments. In the first, we use vUSBf to emulate 10,000 randomly-generated devices. In the second, we use vUSBf to emulate 13 specific configurations identified by the vUSBf authors (after millions of trials) that crash some (older) systems.

On the defensive side, we run Cinch, configured with compliance (§5.2) and logging (§5.5) Policies. If Cinch allows the emulated device to communicate with the blue machine, we account this a failure.

We expect that the overwhelming majority of test cases will not obey the USB specification, and that Cinch's compliance Policy will detect and prevent these cases. As a baseline, we also present the same 10,000 inputs to a system that is not running Cinch.

**Results** are summarized in Figure 4 ("vUSBf"). Cinch's compliance module prevents all emulated devices from connecting to the blue machine. The three most commonly detected violations are: (1) improperly formatted strings, (2) invalid device classes, and (3) invalid or inconsistent number of functions. On the one hand, these results could be argued to be inconclusive because none of these inputs were successful against the baseline setup *without* Cinch. On the other hand, Cinch detected and blocked even the 13 configurations known to crash older systems.

### 7.3 Red team exercise

Our next set of exercises evaluates Cinch against attacks that were not known to us a priori. This is intended to assess Cinch's effectiveness and to avoid some of the bias that may arise when developers choose the attack experiments (as above).

Specifically, we set up a red team that was charged with developing new USB exploits to compromise blue machines; this activity included crafting new vulnerabilities in the blue machine's OS, which was meant to emulate the ongoing process of discovering and patching bugs. In our case, the red team comprised a subset of the authors who were kept separate from the developers of Cinch and worked independently. Interactions between the red team and the developers were tightly controlled, following an evaluation protocol that was documented in advance. Figure 5 summarizes the protocol.

| **Protocol** | There are three rounds, each of which has a *setup*, *match* and *rematch* phase. |
| --- | --- |
| | **Setup:** Red team chooses an OS (which they can modify arbitrarily) and develops exploits that crash the OS. |
| | **Match:** Cinch developers configure Cinch to run the OS provided by the red team as the blue machine; both teams confirm that the exploits crash the OS when Cinch is not present. The Cinch developers deploy Cinch, and the red team mounts its exploits. The Cinch developers collect traces, and both teams document the outcome of the exercise. |
| | **Rematch:** Cinch developers get the traces, and are given the opportunity to analyze and react to them. Then the match phase is rerun. |
| **Attacker knowledge** | **Round 1:** The red team is given access to a technical report that documents an earlier version of Cinch. This models an attacker with limited knowledge of Cinch. |
| | **Round 2:** The red team is given access to a machine that is running Cinch. This models an attacker with black-box access to Cinch, or an attacker that possesses Cinch's binaries. |
| | **Round 3:** The red team is given access to Cinch's source code. This models an attacker with full knowledge of Cinch's logic (but not its configuration). |
| **Developer ability** | Cinch developers freeze Cinch's code prior to the match phase of round 1. After that, Cinch developers may apply configuration-only changes: new signatures, etc. |

FIGURE 5—Summary of the protocol for the red team exercise. This protocol was codified before the exercise began.

**Summary of red team exploits.** The red team developed 3 exploits for Windows and 20 exploits for Linux across the three rounds of the protocol. Some exploits shared the same attack vector but used different payloads.

The Windows exploits attacked a fresh copy of Windows 7; the red team did not install updates because the vulnerabilities their exploits targeted have been patched. Since red team members did not have access or visibility into the Windows USB stack, these exploits were found primarily through fuzzing, guided by past CVEs.

For Linux, the red team installed a modified version of kernel 4.2.0 on a fresh copy of Ubuntu 14.04. In particular, the red team modified a function within HCI that processes USB request blocks (the data structure representing a message in the USB subsystem) to trigger a kernel crash on certain device payloads; introduced a bug in USB core that causes the kernel to crash whenever a device with a certain configuration is connected; inserted a bug in Linux's HID input subsystem (`drivers/input/input.c`) that leads to a null pointer dereference when it receives a specific sequence of input events; and introduced buggy drivers for a USB printer, camera, audio, and HID device.

Finally, the red team noticed that the VFAT filesystem driver in Linux 4.2 does not correctly validate the BIOS Parameter Block (BPB). While they were unable to exploit this bug directly, it can result in an invalid filesystem being mounted. To "enhance" this bug, the red team introduced a null pointer dereference in the BPB handling routine (`fs/fat/inode.c`), triggered by a filesystem with an invalid BPB.

**Results** are summarized in the last 3 sections of Figure 4.

*First round.* The red team developed 7 exploits for this round (2 for Windows and 5 for Linux). In the match phase, Cinch prevented both Windows exploits and 3 out of the 5 Linux exploits. The Windows exploits were prevented by Cinch's architecture rather than by any of its Policies. Specifically, the red machine runs a Linux kernel; that kernel is not vulnerable to either of the Windows exploits and recognizes both connected devices as invalid. As a result, Cinch does not export these devices in the first place, protecting the Windows blue machine.

The two Linux exploits that Cinch was unable to prevent occurred at layers that were outside of its semantic knowledge (VFAT and the input subsystem). Using the traces—collected with Cinch's logging module (§5.5)—the Cinch developers derived signatures. In the rematch phase, these signatures prevented the exploits.

*Second round.* In the match phase, Cinch prevented 14 out of 19 attacks, including attacks from the first round. The rematch phase again relied on signatures; of the remaining five exploits, signatures blocked two. The remaining three succeeded because they are polymorphic: they alter their payload to evade detection.

*Third round.* In the match phase, Cinch prevented 18 out of 23 attacks, including attacks from prior rounds for which signatures were available. In the rematch phase, Cinch was able to defend against an additional exploit using a signature that prevents a particular sequence of key presses from triggering a bug in the modified USB HID driver. The remaining four exploits are polymorphic and escaped evasion by signature and compliance checks.

These results, while preliminary, suggest that Cinch is able to prevent several exploits—primarily those that act as invalid USB devices—without prior configuration; several more can be prevented after deriving signatures. The remaining exploits might be prevented with more intrusive approaches (e.g., sandboxing; §5.3)

**Tradeoff between security and availability.** It is possible to develop more aggressive signatures to prevent polymorphic attacks (for example, using regular expressions); however, this risks disabling benign devices. To ensure that our signatures did not cause such false positives, we established a representative set of benign devices: a USB flash drive, printer, phone, SSD, keyboard, and mouse. After each phase of the experiment, we checked that our signatures did not keep these devices from working.

We found one failure: the signatures for the VFAT exploit prevented the blue machine from communicating with *any* storage device with a VFAT filesystem. We removed the offending signature and accounted that test a failure (i.e., Cinch did not prevent the exploit), since such a signature would not be deployable for most users.

### 7.4 Cinch's flexibility and extensibility

There are two ways that Cinch can currently be extended: through new signatures and configurations to enhance existing Policies (§5), and through new Policies that add new functionality. We discuss our experience in both cases.

**Deriving new signatures.** We take a straightforward approach to deriving signatures for a given attack: we first log malicious traces, and then replay them in a controlled debugging environment. This allows us to analyze the configuration and the attack. We use this information to derive candidate signatures that are on the order of 10–15 lines of JSON; deriving a signature for the exploits in Section 7.3 took roughly 5 to 30 minutes, depending on: (1) the amount of data the exploit sent, and (2) the complexity of the subsystem the exploit targeted.

**Creating new Policies.** Adding a new Policy for Cinch requires implementing an instance of a Rust *trait* [2] (roughly analogous to a Java interface or a C++ abstract class; this trait is defined in the Gateway library, §6.2) that processes USB transfers, and adding the new Policy to Cinch's configuration file. Based on this configuration, Cinch's module subsystem automatically dispatches USB transfers to configured chains (§6.2). To give an idea of Policies' complexity, Cinch's largest—compliance—is 2500 SLoC while the rest average just 180 SLoC.

### 7.5 What are the costs of Cinch?

To understand the performance cost associated with using Cinch, we investigate two microbenchmarks, one for latency and one for throughput. We use Debian Jessie (Linux 4.2.0) as the blue machine's OS.

**Is Cinch's added latency acceptable?** To quantify the delay introduced by the components of Cinch, we connect the blue machine and another machine on a local network, using an Ethernet-over-USB adapter. We record the round-trip time between the two machines (using ping) as we add components of Cinch. Figure 6 shows the results.

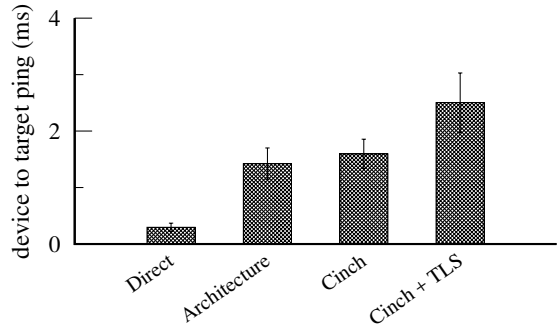

FIGURE 6—Round-trip time between the blue machine and USB device as components of Cinch are progressively added. Results are averaged over 1000 pings and error bars represent one standard deviation of the mean.

|                          | direct     | Cinch      |
|--------------------------|------------|------------|
| **USB 2 device** (flash drive) |            |            |
| % of CPU cycles          | 1.8 %      | 8.1%       |
| memory                   | 9 MB       | 205 MB     |
| I/O throughput           | 181.6 Mbps | 145.6 Mbps |
| Encrypted I/O throughput | –          | 35.4 Mbps  |
| **USB 3 device** (SSD)   |            |            |
| % of CPU cycles          | 5.6%       | 38.2%      |
| memory                   | 9 MB       | 207 MB     |
| I/O throughput           | 3.4 Gbps   | 2.1 Gbps   |

FIGURE 7—Resource consumption of Cinch when transferring a 1 GB file from storage devices to the blue machine. The "direct" baseline is a setup where devices are connected directly to the blue machine. Entries are the mean over 20 trials; standard deviation is less than 5%. We do not report encrypted throughput for the SSD because the crypto adapter does not support USB 3.

For our baseline, we connect the Ethernet-over-USB adapter directly to a USB port on the host (Fig. 6, "Direct"). We next attach the device to the red machine and export it to the blue machine through the Tunnel without the Gateway (i.e., the Tunnel runs directly to the blue machine); this arrangement demonstrates the latency cost of Cinch's use of virtualization (Fig. 6, "Architecture"). Next, we add the Gateway to the above configuration, enabling all of Cinch's Policies (§5), demonstrating the overhead when the Gateway interposes on all USB transfers (Fig. 6, "Cinch"). Finally, we place the crypto adapter (§5.4) in between the Ethernet-over-USB device and the Gateway (Fig. 6, "Cinch + TLS").

Each component of Cinch adds moderate delay, with the full setup (including the crypto adapter) resulting in a round-trip time of less than 2.5 ms. We believe that this delay is acceptable for latency-sensitive input devices; as a comparison, high-performance mechanical keyboards introduce delays on the order of 5 ms between successive keystrokes (for debouncing [69, 107]).

**What is Cinch's impact on throughput and other resources?** We read 1 GB of data from a USB storage device to the blue machine and measure the throughput, memory consumption, and CPU load with and without Cinch; we repeat these experiments 20 times. Storage devices range in performance, so we experiment with two: a USB 2 flash drive and a USB 3 SSD.

Figure 7 tabulates the results. For the flash drive, Cinch achieves $0.8\times$ the baseline's throughput. There are two main reasons for this: (1) Cinch copies USB transfers at several stages in its architecture; and (2) USB 2 flash drives use exclusively synchronous transfers, meaning that Cinch's added latency translates to lower throughput. For the USB 3 SSD, Cinch achieves $0.6\times$ the baseline's throughput. Unlike in USB 2, USB 3 storage devices use asynchronous transfers and allow multiple in-flight requests. The primary overhead is thus memory copies.

With regard to CPU and memory use, Cinch has modest overhead. The memory Cinch consumes, which is primarily allocated to running the red machine, is in line with the cost of other security applications (e.g., antivirus).

## 7.6 Summary and critique

Our evaluation shows that Cinch can prevent previously documented vulnerabilities, fuzzing attempts, and crafted attacks, even without attack-specific configuration. Augmented with a signature database, its success is even higher, though none of its Policies are well suited to defeating polymorphic attacks. In this respect, Cinch is comparable to related tools in network security: it rules out certain classes of vulnerabilities and can be adapted to address specific issues, but it is not perfect. Cinch's extensibility also seems reasonable, though our metrics here are subjective; and the performance impact, while not negligible, may be a good trade-off.

While this evaluation suggests that Cinch is a step in the right direction, it is far from definitive. First, we have likely not explored the full attack space, especially with regard to attacks on the non-USB portions of the kernel and on user software. Second, the red team comprised authors rather than disinterested parties, which may bias the security evaluation. Third, most systems are considered usable by their implementers; a neutral, non-expert operator may have a different perspective. Finally, Cinch's performance impact may be acceptable for a wide range of devices, but others (e.g., audio and video devices) have more stringent latency requirements that Cinch might not meet, especially when using the crypto adapter.

## 8 Related work

Cinch's contribution is architectural: most of its mechanisms are adapted from prior works and existing areas of research. Nevertheless, we are not aware of any other system that addresses the full space of attacks described in Section 3.

**USB security mechanisms (similar problem, different mechanisms).** One can purchase an adapter that prevents data interchange on the USB bus, converting the bus into power lines only [51]. A software version of this protection is a set of Linux kernel patches known as grsecurity [23], which essentially disable hotplug. This "air gap ethos"—provide defense by eliminating connectivity—conflicts with Cinch's aim of controlled interaction.

Qubes [45] is a distribution of Linux that makes extensive use of virtualization to create isolated privilege domains for *applications*. Qubes can place USB devices in their own virtual machines (USB VMs). A device's transfers are delivered to its USB VM, and hence applications accessing that device need to live on that VM, wherein the threats enumerated in Section 3 are reprised. An exception is that Qubes allows a user to safely share USB storage devices from a USB VM with other VMs on the system by exporting them as block devices. Qubes also supports exporting keyboards and mice from a USB VM, but its developers warn that doing so risks exposing the system to attacks [62].

The udev user space daemon on Linux [56, 110] implements finer-grained policies than Qubes, akin to Cinch's containment Policy (§5.3). However, udev can itself be attacked: udev requires the kernel to interact with every device that connects, so the device has an opportunity to attack the host machine before udev makes a policy decision. There are many commercial offerings that enable access control for USB devices [13, 15, 18, 22, 33, 35, 37, 48, 50, 55]; the issues with these are similar to udev.

USBFILTER [147] enables more precise and expressive access control policies than udev. Furthermore, these policies are enforced throughout the lifetime of the interaction rather than only at connection time. In particular, a user can define rules to dictate which entities (processes and drivers) can interact with a device (and vice versa). This is similar to Cinch's containment Policy (§5.3), but USBFILTER's rules support finer-grained statements, for example, restricting interaction to particular processes. The tradeoff is that it requires instrumenting the host's OS to trace USB transfers all the way to the requesting processes and drivers. USBSec [149] brings a similar tradeoff: it extends the USB protocol with mutual authentication between the host and a compatible device (providing a subset of Cinch's crypto Policy functionality; §5.4) but requires changes to the host's USB stack.

GoodUSB [146] loads devices in a sandboxed environment and prompts the user to enable functions based on a device's claimed identity. This is similar to (but richer than) Cinch's containment Policy (§5.3), which could be enhanced accordingly. GoodUSB's mechanisms might also be used to bootstrap Cinch's crypto overlay, as mentioned in Section 5.4.

Under UScramBle [124], devices provide a key to the host that can be used to encrypt further messages; the message goes upstream and thus is not broadcast across the bus (§3.3). This prevents eavesdropping for USB 2 and earlier, but unlike Cinch's crypto overlay (§5.4), it cannot protect against malicious or compromised hubs that see the key.

Of the foregoing, only USBFILTER, USBSec, and GoodUSB address masquerading attacks (with the help of the user; §5.3); eavesdropping (§3.3) is out of scope for these systems. In contrast, UScramBle addresses eavesdropping but not masquerading.

**Device driver isolation and reliability (complementary problem, overlapping mechanisms).** There is a vast literature on device driver containment and reliability. We will go over some of it, but we can only scratch the surface (a helpful survey appears in SUD [80]). We note at the outset that Cinch borrows mechanisms from many of these works: placing drivers in a separate virtual machine [93, 95, 114], isolating a device with the IOMMU [105], and leveraging hardware-assisted I/O virtualization [105, 114, 145]. However, the threat and the resulting architecture are different.

Specifically, work that isolates faulty device drivers [80, 83, 93, 95, 96, 105, 112, 114, 127, 143–145, 152] assumes that hardware obeys its specification (and, with the exception of SUD [80], that drivers may be buggy, but not malicious). The same assumption about hardware is made by work that validates the commands passed to devices [152], eliminates bugs from drivers [130], and synthesizes drivers that are correct by construction [131, 132]. There is work that aims at tolerating hardware faults [108], but these faults are non-malicious and constrained (for example, flipped bits) compared to the types of attacks outlined in Section 3.

As a result of the assumption about faithful hardware, masquerading and eavesdropping are out of scope; often, devices that deviate from specification (§3.3) are, too. On the other hand, Cinch does not provide comprehensive protection against compromised drivers (though it can sanitize drivers' inputs, as outlined in §5.2). For this reason, the works covered above are complementary to—and in many cases composable with—Cinch.

**Secure peripheral interaction (different problem, overlapping mechanisms).** Kells [82], USB Fingerprinting [76, 113], and work by Wang and Stavrou [150] allow a USB device to establish the identity of a host. The first two works are defense mechanisms *against* the host: they prevent compromised OSes from corrupting devices or propagating malware; the latter is an attack primitive and allows a malicious device to compromise hosts selectively. Cinch's crypto overlay (§5.4) also allows a device to identify a host (since connections can be mutually authenticated; §6.3), but the goal is to prevent eavesdropping and device masquerading.

SeRPEnT [151] and Bumpy [121] provide a safe pathway from devices, through an untrusted host machine, to a trusted, remote machine. SeRPEnT provides a similar abstraction to Cinch's crypto overlay (§5.4), and its mechanism is comparable to Cinch's crypto adapter. Bumpy's goal, however, is remote attestation of user input rather than prevention of masquerading attacks; its mechanisms are based on trusted hardware. Both of these works target wide area networking, while Cinch focuses on intra-host communication.

Zhou et al. [155] allow trusted applications running on top of untrusted OSes to securely communicate with I/O devices. This is done via a trusted hypervisor that mediates access to hardware by both the trusted and untrusted components. Cinch also interacts with peripheral devices via an untrusted intermediary, but the architecture, mechanisms, goals, and threat model are all different.

**Separation kernels and network security (related problems, related mechanisms).** Two other research areas deserve special mention. The first is Rushby's separation kernel [129], in which the operating system is architected to make a computer's components interact as if they were part of a distributed system (see [81] and [122] for modern implementations). The foundational observation of this work—that networks are a useful abstraction for interposition—is one that we share. However, our goals and scenario are different. The separation kernel was intended to be a small kernel, with compartmentalized units that could be formally verified, and it provided separation through information flow control. In contrast, our scenario is commodity operating systems, and we are seeking to apply the conceptual framework of network security.

This brings us to network security itself. Cinch owes a substantial debt to this field, borrowing as it does concepts like firewalls, deep packet inspection, and virtual private networks. Moreover, the recent trend toward Network Function Virtualization (NFV) [119, 138] applies I/O virtualization (as do Cinch and some of the works cited earlier), but the point in NFV is to make middleboxes virtual, for reasons of configurability and cost.

## 9 Summary and conclusion

Cinch was motivated in large part by the observation that hardware security is recapitulating the history of network security. Originally, the Internet was a comparatively small number of mutually trusting organizations and users. As a consequence, there was relatively little focus on support for security within the network infras-tructure. With the explosion of Internet users, spurred by changing economics, security suddenly became a serious problem. Similarly, commodity operating systems have relatively few safeguards against misbehaving hardware, reflecting a time when peripheral devices could be trusted. But, with the rapid decline in the barriers to producing plug and play peripherals, those days have come to an end—and Cinch aims to be useful in the world ahead.

Although Cinch's individual mechanisms have ample precedent in the literature, the architecture and the synthesis is novel, to the best of our knowledge. Moreover, as the evaluation results make clear, the implementation is pragmatic and surprisingly powerful. Looking at this fact, we feel comfortable stating that we have identified a good abstraction for the problem at hand.

To be clear, we are not saying that Cinch uniquely enables any one piece of its functionality (§5); rather, the abstraction makes it natural to develop and deploy what would require far more work under alternative solutions (§8).

We are also not saying that Cinch is comprehensive. Indeed, besides the limitations covered earlier (§1, §4.2, §7.6), some of Cinch's solutions are effective only with additional mechanisms. As a key example, providing authentication and privacy with Cinch requires certificates or pairing, and device modifications. However, certificates are compatible with the chain of trust inherent in purchasing hardware, pairing is similar to the permissions model on mobile devices, and the required modifications are not onerous, as our implementation of the adapter (§6.3) indicates. As another example, Cinch's compliance Policy (§5.2) would be strengthened by formal verification.

Despite the issues, Cinch appears to improve on the status quo. Of course, it is possible that, if Cinch were widely deployed, it would only escalate an arms race, and drive attackers to find ever more esoteric vulnerabilities. On the other hand, security is always about building higher fences, and the considerations at the heart of our work could guide the future design of peripheral buses and drivers.

## Acknowledgements

This paper was aided by conversations with Andrew Baumann, Adam Belay, Sergio Benitez, Kevin Butler, Christian Huitema, Trammell Hudson, Ant Rowstron, Dennis Shasha, Jeremy Stribling, Ymir Vigfusson, and Junfeng Yang; and substantially improved by the detailed comments of the SOSP and USENIX Security reviewers. This work was supported by NSF grants CNS-1055057, CNS-1423249, and CNS-1514422; AFOSR grant FA9550-15-1-0302; and ONR grant N00014-14-1-0469.